\documentclass[aps,twocolumn]{revtex4}

\usepackage{epsfig}
\usepackage{psfrag}
\newcommand{\be}{\begin{equation}}
\newcommand{\ee}{\end{equation}}
\newcommand{\ba}{\begin{array}}
\newcommand{\ea}{\end{array}}

\begin{document}

\title{Polymers Confined between Two Parallel Plane Walls}

\author{Hsiao-Ping Hsu and Peter Grassberger}

\affiliation{John-von-Neumann Institute for Computing, Forschungszentrum J\"ulich,
D-52425 J\"ulich, Germany}

\date{\today}

\begin{abstract}
Single three dimensional polymers confined to a slab, i.e. to the region between 
two parallel plane walls, are studied by Monte Carlo simulations. They are 
described by $N$-step walks on a simple cubic lattice confined to the region 
$1 \le z \le D$. The simulations cover both regions $D << R_F$ and $D >> R_F$
(where $R_F \sim N^\nu$ is the Flory radius, with $\nu \approx 0.587$),
as well as the cross-over region in between. Chain lengths are up to $N=80,000$,
slab widths up to $D=120$. In order to test the analysis program and to check
for finite size corrections, we actually studied three different models:
(a) Ordinary random walks (mimicking $\Theta$-polymers); (b) Self-avoiding 
walks (SAW); and (c) Domb-Joyce walks with the self-repulsion tuned to the point 
where finite size corrections for free (unrestricted) chains are minimal.
For the simulations 
we employ the pruned-enriched-Rosenbluth method (PERM) with Markovian 
anticipation. In addition to the partition sum (which gives us a direct 
estimate of the forces exerted onto the walls), we measure the density 
profiles of monomers and of end points transverse to the slab, and the 
radial extent of the chain parallel to the walls. All scaling laws and 
some of the universal amplitude ratios are compared to theoretical predictions.

\end{abstract}

\maketitle
\section{Introduction}

Although the behaviour of flexible polymers in a good solvent
confined to different geometries has been studied for many 
years~\cite{deGen,Eisenriegler93}, there are still a number
of open questions. In the present work we shall only
discuss single polymer chains between two parallel walls which act 
only as geometric constraints, without any
energetic effects. 

Theoretically, this problem is rather well understood. All important 
scaling laws have been formulated, including the cross-over from the 
region of narrow slabs (where the distance $D$ between the walls is 
smaller than the Flory diameter of a free coil) to the opposite case 
of wide slabs. In particular, there exists an important theoretical 
prediction: Near such a wall the monomer density profile increases as 
\be
   \rho(z) \sim z^{1/\nu}\;,     \label{rhoz}
\ee
where $z$ is the distance from the wall and $\nu$ is the Flory 
exponent~\cite{deGen}. This is supposed to hold for all dimensions (not only 
for $d=3$), and both for ordinary random walks (ideal polymers) for 
which $\nu=1/2$ and for self avoiding walks with $\nu\approx 0.587$
(in $d=3$).

It is intuitively obvious that the force exerted by the 
polymer onto the wall is proportional to the monomer density near the wall. 
The ratio between the two can be expressed in terms of a universal amplitude 
ratio which is easy to calculate for ideal chains, and which was calculated
by Eisenriegler~\cite{Eisenriegler} as an expansion in $\epsilon=4-d$.
Several authors have tried to verify these detailed predictions by Monte Carlo
simulations~\cite{Webman,Ishinabe, Milchev,Joannis}, but the results are 
not yet convincing. While the scaling of the density near the wall is 
roughly verified, the amplitude ratio consistently has come out too large,
casting even doubt on the validity of the $\epsilon$-expansion.

In Ref.~\cite{hg03}
we had studied confined polymers in a strip in two 
dimensions where the amplitude ratio had been predicted 
by Eisenriegler~\cite{Eisenriegler2} 
(using {\it conformal invariance} results of
Cardy et al.~\cite{cardy}).
There we verified all predictions, but we found that this was less easy than 
anticipated: There are very large corrections to Eq.~(\ref{rhoz}) which can
easily be missed, and overlooking them would give wrong estimates of the 
amplitude ratio. This suggests of course that the same effect was the 
source of difficulties in $d=3$.

It is the purpose of the present paper to present a careful numerical
study, in order to settle these questions. We not only simulated much
larger systems than previous authors, going to chain lengths up to 
$N=80,000$, slab widths up to $D=120$, and collecting rather high 
statistics. Since it is well known that asymptotic scaling of unconstrained 
self-avoiding walks
(SAWs) is reached rather slowly, with correction terms decreasing only 
as $N^{-0.5}$~\cite{LiMadrasSokal,belohorec,gss}, we studied also the 
Domb-Joyce (DJ) model~\cite{dj} with $w=0.6$ (where convergence to 
asymptotia is much faster~\cite{belohorec,gss}) in addition to SAWs. 

The DJ model is defined by the partition sum 
\be
    Z_N(w) =\sum_{configs.} w^{\kappa}    \label{ZDJ}
\ee
where the sum extends over all random walk (RW) configurations with $N$ 
steps, $0 \le w \le 1$, and $\kappa$ is the total number of monomer pairs 
occupying a common site. For $w=1$ the DJ model describes just ordinary
random walks. For $w=0$ it is just the SAW model. Asymptotically (for 
$N\to\infty$) the model is in the SAW universality class for all $w<1$, 
but the speed with which the renormalization group fixed point is approached 
depends on $w$. Moreover, it is approached from opposite sides when $w<w^*$
and when $w>w^*$, with $w^*\approx 0.6$~\cite{gss,belohorec}. For $w=w^*$
the approach to asymptotia is fastest, and we thus expect also smaller 
finite size corrections for the present problem of confined polymers.

Finally, we also performed simulations of the ordinary RW model, just to 
check the simulation and analysis programs, as everything can be calculated
there analytically.

For the simulation we used the pruned-enriched-Rosenbluth method 
(PERM)~\cite{g97} with $k$-step Markovian anticipation~\cite{Frauenkron98,
Frauenkron99,Caracciolo,hg03}. Apart from being fast (notice that the 
pivot algorithm which is very fast for unconstrained polymers~\cite{LiMadrasSokal}
is very inefficient for narrow slabs), it has the advantage that the partition
sum is computed by default with very high precision. Thus we could estimate
the dependence of the monomer fugacity on the width $D$, and from that 
directly the total forces exerted onto the walls. For more details see 
Ref.~\cite{hg03}.

Details of the scaling predictions are discussed in Sec. II, 
while results and their comparison with theoretical predictions are 
presented in Sec.~III. Conclusions are finally given in Sec.~IV.

\section{Scaling predictions} 

The end-to-end distance $R_N$ of a free SAW in infinite $d-$dimensional volume
scales as 
\be
   R_N^2 \equiv \langle ({\bf x}_N-{\bf x}_0)^2\rangle \approx 
       d\;(kN)^{2\nu_d}\;(1+b/N^{\Delta_d})  \label{R2}
\ee
where $\nu_d$ is the Flory exponent, $\Delta_d$ is the leading correction 
to scaling exponent, and $b$ and $k$ are non-universal constants
which depend on the microscopic realization. In $d=2$ one has $\nu_2=3/4$
~\cite{deGen}, while the best published estimates for $d=3$ are 
$\nu_3 = 0.5877(6)$~\cite{LiMadrasSokal},
$0.5874(2)$~\cite{prellberg}, and $0.58758(7)$~\cite{belohorec}. 
In the latter paper it was assumed that the leading correction to Eq.~(\ref{R2})
is $1+b/N^{0.5}$. Since this is questionable (our own simulations gave 
$\Delta_3\approx 0.45$~\cite{gss,unpub}), we use in the following 
$\nu_3 = 0.58765(20)$ which also incorporates results from extensive 
simulations of the DJ model with $w=0.6$~\cite{unpub}. For the DJ model
$w^*$ is defined by $b(w=w^*)=0$. The absence of large corrections to scaling 
leads to a rather precise estimate of the constant $k$ in case of the DJ model
with $w=0.6$: $k=0.3259(4)$~\cite{unpub}, but for 3-d SAWs the estimate is 
much less stable. The value $k=.4640(4)$ of 
Ref.~\cite{LiMadrasSokal} depends crucially
on the estimate $\Delta\approx 0.56$ made by these authors. Assuming instead
$\Delta=0.5$ as in Ref.~\cite{belohorec}
and the value of $\nu_3$ found by these 
authors, the same data would give $k\approx 0.4655$, while an even larger value 
would be obtained if $\Delta<0.5$. In the following 
we shall use $k=0.4657(7)$ for 3-d SAWs.

Equation~(\ref{R2}) with $d=2$ also describes the behaviour of the {\it parallel}
components (i.e. parallel to the wall) in the regime $1\ll D \ll N^{\nu_3}$
where the polymer is essentially two-dimensional. The constant $k$ depends 
then on the slab width $D$. A scaling ansatz for 
the cross-over between the two regimes $1\ll N^{\nu_3}\ll D$ and 
$1\ll D \ll N^{\nu_3}$ is~\cite{deGen} 
\be
      R^2_{N,\;||}(D) = R_N^2\;\Phi(R_N/D)   \label{R2par}
\ee
where $R_N^2$ is given by Eq.~(\ref{R2}) with $d=3$, and
\be
   \Phi(\eta)=\left\{\begin{array}{ll}
      2/3 & \;\;{\rm for}\;\;\, \eta \rightarrow 0 \;, \\
      \eta^{2(\nu_2/\nu_3-1)} & \;\;{\rm for}\;\;\, \eta \rightarrow \infty \; .
      \end{array}
      \right . \label{Phi}
\ee
It leads to the prediction~\cite{Milchev} 
\be
   k(D) \sim D^{(\nu_3-\nu_2)/\nu_2\nu_3}.
\ee

The partition sum of a free SAW in infinite volume scales for $N\to\infty$
as 
\be
   Z_N = \mu_\infty^{-N} N^{\gamma_d-1}\;const    \label{ZN}
\ee
with $\mu_\infty$ being the critical fugacity per monomer, and with $\gamma_d$
being a universal exponent. In two dimensions $\gamma_2=43/32$~\cite{deGen}, 
while the best published estimate for $d=3$ is $\gamma_3 = 1.1575(6)$~\cite{Causo}. 
In the following we shall use the estimate $\gamma_3 = 1.1575(3)$ obtained 
from the DJ model with $w=0.6$~\cite{unpub}. For 3-d SAWs one has $\mu_\infty=0.213491(4)$
from exact enumerations~\cite{MacD} and $\mu_\infty = 0.2134910(3)$ from Monte Carlo 
simulations~\cite{gss,unpub}. In the following we shall use the latter. For the 
DJ model we use $\mu_\infty = 0.18812145(7)$~\cite{unpub}.

Again we must expect that the same ansatz, with $\gamma_3$ replaced by 
$\gamma_2$, with $\mu_\infty$ replaced by $\mu(D)$, and with the constant 
replaced by $c(D)$, holds for slabs in the limit $D\ll N^{\nu_3}$:
\be
   Z_N(D) = \mu(D)^{-N} N^{\gamma_2-1}\;c(D) \quad {\rm for}\;\; D\ll N^{\nu_3}\;.
      \label{ZNDinf}
\ee
The corresponding cross-over ansatz is then
\be
   Z_N(D)=Z_N\;\Psi(R_N/D)                  \label{ZND1}
\ee
with
\be
 \Psi(\eta)=\left\{\begin{array}{ll}
      const & \;\;{\rm for}\;\; \, \eta \to 0 \; ,\\
      \eta^{(\gamma_2-\gamma_3)/\nu_3} \exp(-{a\over {\mu_\infty}k}
(\eta/\sqrt{3})^{1/\nu_3})  &
     \;\;{\rm for} \;\;\, \eta \to \infty \;,
     \end{array}
    \right . \label{PHI}
\ee
\be
   \mu(D) = \mu_\infty + a D^{-1/\nu_3}\;,   \label{MUD}
\ee
and 
\be
   c(D) = const \; D^{(\gamma_2-\gamma_3)/\nu_3}\;.
\ee
Notice in particular that the $D$-dependence of $\mu(D)$, Eq.~(\ref{MUD}),
follows directly from the scaling ansatz for the cross-over.
We should also point out that the partition function for a polymer in a slab
is defined as the sum over all walks starting at fixed ${\bf x}_{0,\|} = (x_0,y_0)$,
but averaged over all $z_0\in [1,D]$.

The force exerted onto the wall is most straightforwardly expressed
in terms of the work done when moving one of the walls, i.e. by the
dependence of the free energy -- and thus also of the partition sum --
on $D$,
\be
   F = k_BT\; {d\ln Z_N(D) \over dD}\;,        \label{F}
\ee
where we have introduced a dummy temperature $T$ which can take any
positive value. From Eqs.~(\ref{ZNDinf}) and (\ref{MUD}),
the force per monomer is then obtained as
\be
   f = F/N = k_BT \; {a\over \nu_3\mu_\infty}\; D^{-1-1/\nu_3}        \label{ff}
\ee
in the limit of $D \rightarrow \infty$ and $N>>D^{1/\nu_3}$.

The monomer density near the wall is predicted to scale with Eq.~(\ref{rhoz}).
For ordinary random walks this gives $\rho(z)\sim z^2$, but in that 
case one can compute $\rho(z)$ exactly, with the result 
\be
   D\rho(z) = 1-\cos(2\pi z/D) \equiv f(z/D)      \label{rho-rw}
\ee 
for $D\gg 1$ (we normalize $\rho(z)$ such that $\sum_{z=1}^D 
\rho(z)=1$). For walks with excluded volume $\rho(z)$
is not known, but one expects $D\rho(z)$ to be universal.

One should expect that the density near the walls is 
proportional to the force per monomer.
Indeed it was shown by Eisenriegler~\cite{Eisenriegler} that
\be
     \lim_{z \rightarrow 0} k \frac{\rho(z)}{z^{1/\nu_3}}=
     B \frac{f}{k_BT} = B {a\over \nu_3\mu_\infty}\; D^{-1-1/\nu_3}    \label{rhof}
\ee
with $B$ being a universal amplitude ratio. For ideal chains one
has $B=2$, while for chains with excluded volume in $4-\epsilon$
dimensions one has $B \approx 2(1-b_1 \epsilon)$
with $b_1=0.075$~\cite{Eisenriegler}. In three dimensions this gives 
the prediction $B \approx 1.85$.
   
Finally, the density profile of end points scales near the walls with
a new exponent which is related to the surface exponent in spin 
systems~\cite{diehl}. For polymers, the standard way to introduce 
this exponent is via the partition sum of a SAW, one end of which is 
glued to an impenetrable wall. For this system one has 
\be
   Z^{(1)}_N \sim \mu_\infty^{-N} N^{\gamma_d^{(1)}-1}
\ee
with $\gamma_3^{(1)}=0.679(2)$~\cite{gh94}. The end point density 
then scales like~\cite{ekb82} 
\be
   \rho_{end}(z) \sim z^{(\gamma-\gamma^{(1)})/\nu} \sim z^{0.814(6)}
      \label{rhoend}
\ee
where the right hand side holds for $d=3$.

\section{Results}

\subsection{Random walks}

In order to test our simulation and analysis methods,
we first simulate the simple model of RWs on a simple cubic 
lattice between two hard walls at $z=0$ and $z=D+1$, confining
the polymer to $1\le z \le D$.
We simulated width $D$ up to $80$ and chain length between
$3500$ (for $D=4$) and $80000$ (for $D=80$). 

\begin{figure}
  \begin{center}
\mbox{(a)}
\psfig{file=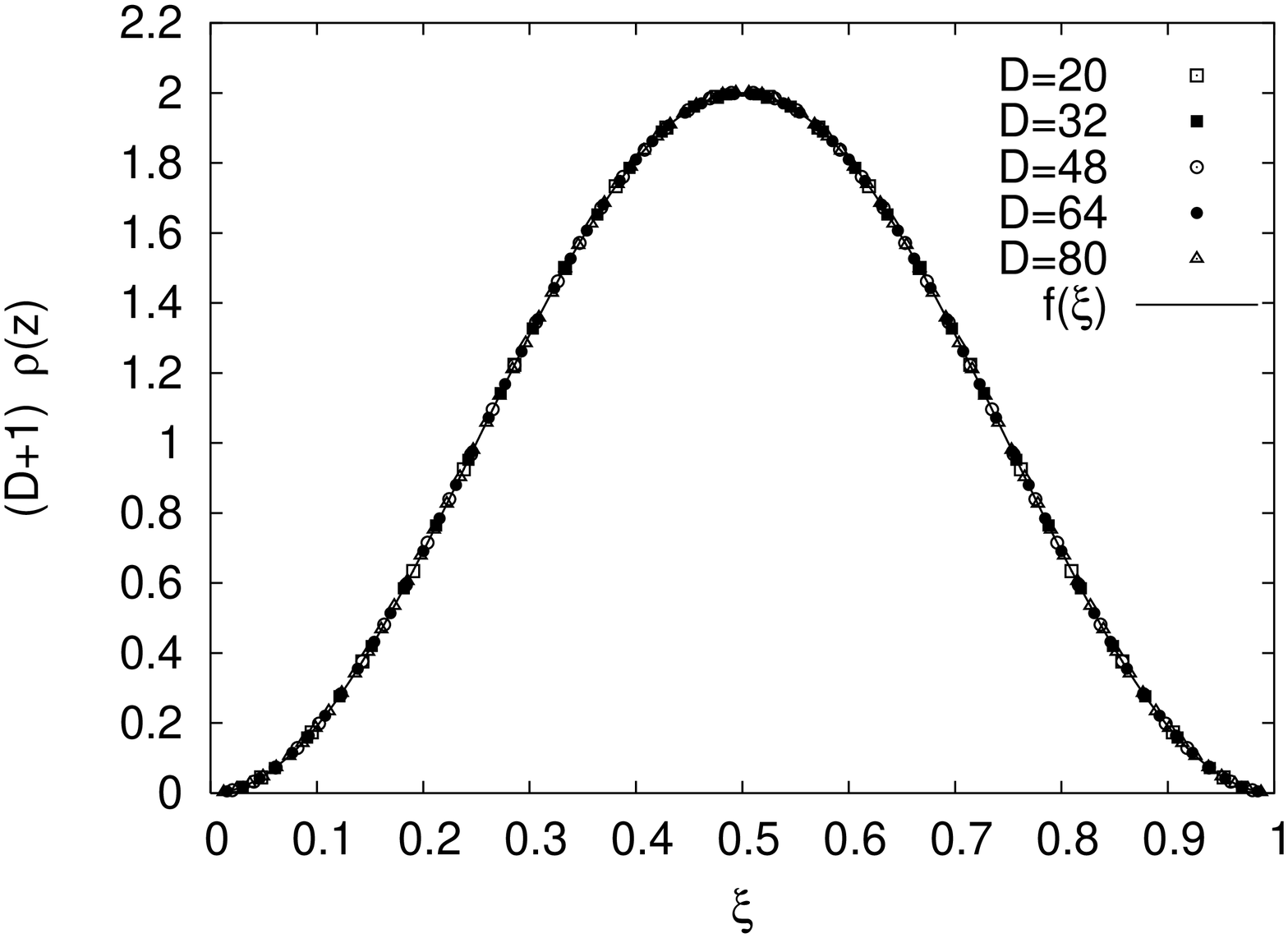,width=5.0cm, angle=270} \\
\mbox{(b)}
\psfig{file=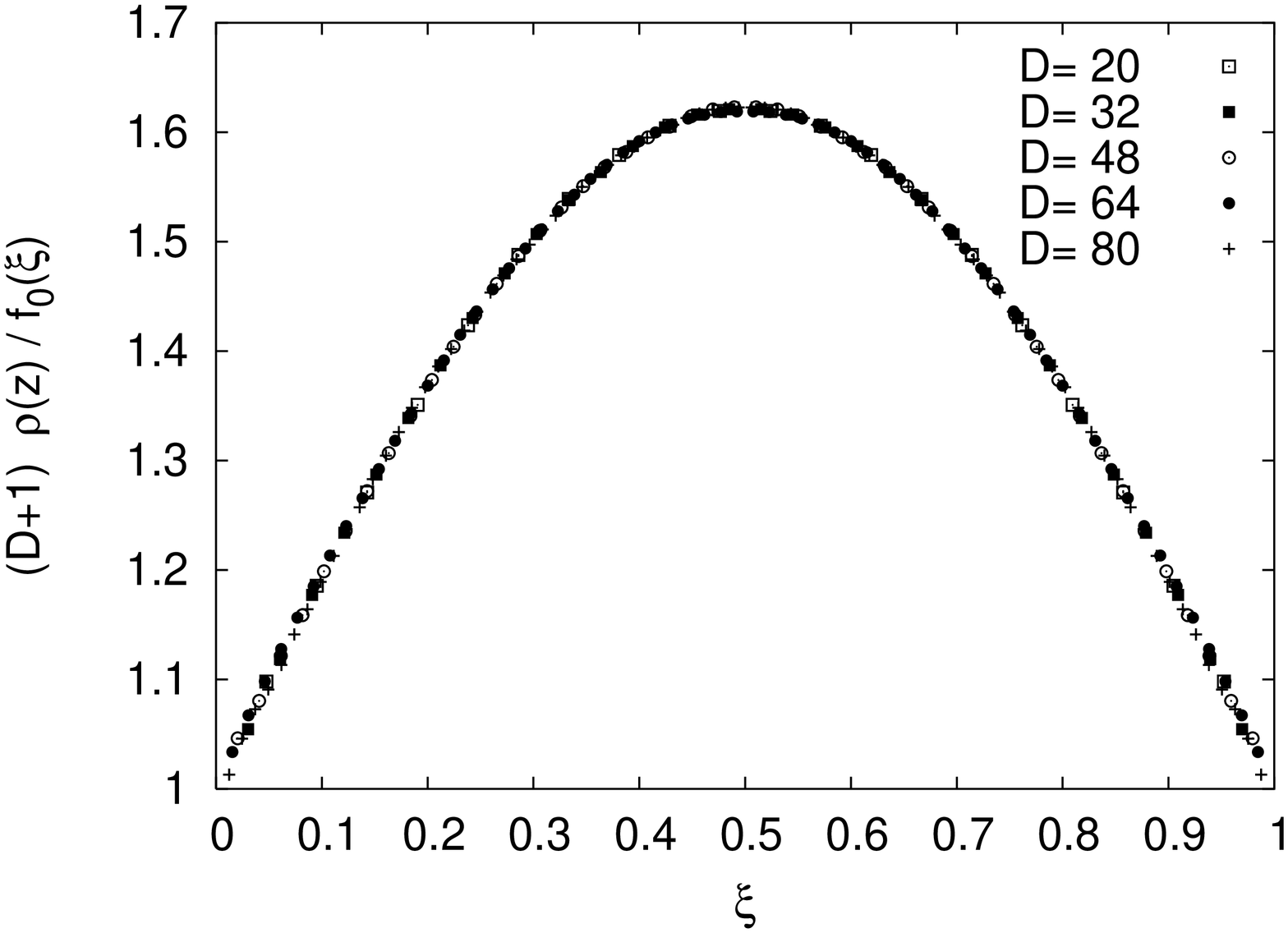,width=5.0cm, angle=270}
   \caption{Rescaled values of the monomer density, $(D+1)\;
      \rho(z)$ against $\xi=z/(D+1)$ for ordinary random walks.
      Also plotted is the function $f(\xi) = 2\sin^2(\pi \xi)$.
     (b) The same values, but divided by $f_0(\xi)=2\pi^2[\xi(1-\xi)]^2$.}
    \label{rw-rhoz}
  \end{center}
\end{figure}

Monomer densities are shown in Fig.~\ref{rw-rhoz}. They were obtained by 
averaging over the central part of the chains, excluding 10 \% on
either side to avoid errors from the fact that Eq.~(\ref{rhoz}) should
hold only far away from the ends, for monomer indices $n$ satisfying 
$D^2\ll n \ll N-D^2$ (we should mention that $N/D^2>10$ for all data sets).
For finite $D$ the scaling has to be slightly modified, by replacing 
in Eq.~(\ref{rho-rw}) $z/D$ by $\xi = z/(D+1)$ and $D\rho(z)$ by 
$(D+1)\rho(z)$. We see that all data, even for small $D$, fall precisely
onto the predicted curve. To show that also the regions near the walls
are correctly sampled, we plot in panel (b) of Fig.~\ref{rw-rhoz} the 
same data but divided by the product of the two power laws for $z\approx 0$
and $z\approx D$, $(D+1)\rho(z) / f_0(\xi)$ with $f_0(\xi)=2\pi^2[\xi(1-\xi)]^2$.

\begin{figure}
  \begin{center}
\psfig{file=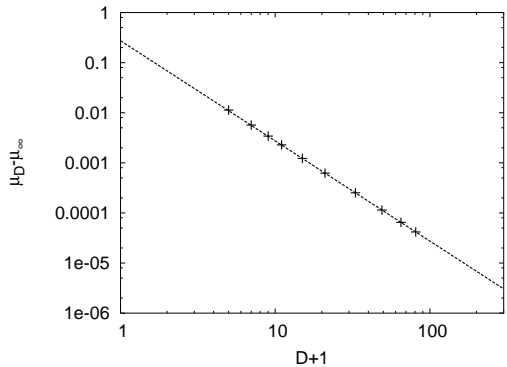,width=5.0cm, angle=270}
   \caption{Log-log plot of $\mu_D-\mu_\infty$ against $D+1$, for ordinary
      random walks. The dashed line is $\mu_D-\mu_\infty = 0.2741/(D+1)^2$
      and gives our best "extrapolation'' of the data for large $D$.}
    \label{rw-mu}
  \end{center}
\end{figure}

Critical fugacities are determined by plotting $Z_N \mu_D^N$ against
$\log N$ and demanding that these curves become horizontal for large $N$.
Results are shown in Fig.~\ref{rw-mu},
where we plot $\mu_D-\mu_\infty$ with $\mu_\infty=1/6$. 
The dashed line is not a fit to the data, but fits their extrapolation 
to $D\to\infty$, $\mu_D-\mu_\infty = 0.2741/(D+1)^2$. This agrees with
Eq.~(\ref{MUD}) (since $\nu=1/2$) and gives $a=0.2741(2)$, where the error
is obtained by assuming that the slope is 2 as predicted. On the other 
hand, Eqs.~(\ref{rhof}) and (\ref{rho-rw}) together with $B=2,\; k=1/3$, 
and $\mu_\infty=1/6$ give $a=\pi^2/36 = 0.27416$, in perfect agreement.
We should point out that Eq.~(\ref{MUD}) is significantly violated for 
small $D$ in this model, and becomes exact only for large $D$. We shall see 
the same behaviour also for SAWs and for the DJ model, and we had seen the 
same also in $d=2$~\cite{hg03}.

\begin{figure}
  \begin{center}
\psfig{file=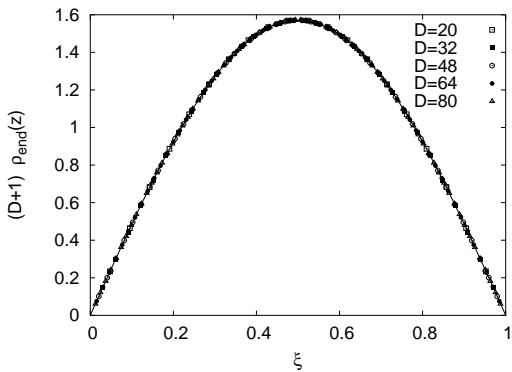,width=5.0cm, angle=270}
   \caption{Rescaled values of the probability $\rho_{end}(z)$ that the chain
       end is at the distance $z$ from a wall against $\xi=z/(D+1)$, for
       ordinary random walks.
       The solid line is the function ${\pi\over 2} \sin(\xi)$.}
            \label{rw-rhoend}
  \end{center}
\end{figure}

In Fig.~\ref{rw-rhoend} we show the transverse distribution of chain ends.
For ordinary RWs it should be just the square root of the monomer density,
i.e. proportional to $\sin(\pi \xi)$. This is obviously true for our 
data, with high precision.

Finally we should mention that the cross-over ansatzes, Eqs.~(\ref{R2par}) 
and (\ref{ZND1}) become trivial for RWs, since the critical exponents are 
the same in $d=2$ and $d=3$. We therefore also do not show data for 
$R_{N,\|}$ which can be calculated approximately by assuming that all 
steps are uncorrelated, and that vertical steps occur with probability 
1/3 inside the slab and with probability 1/5 at the boundaries.

\subsection{Self-avoiding walks}

Let us first discuss the trivial case $D=1$. In this case we have ordinary
2-d SAWs, and therefore we can use the comparison with the known results
as a test for our algorithm. Our simulations, with $N=3000$, gave 
indeed perfect agreement for the critical exponents, and also the 
amplitude for the end-to-end distance, 
$R^2_{||}/N^{2\nu_2}\approx0.771(1)$, in agreement with the value obtained
in~\cite{LiMadrasSokal}.

\begin{figure}
  \begin{center}
\psfig{file=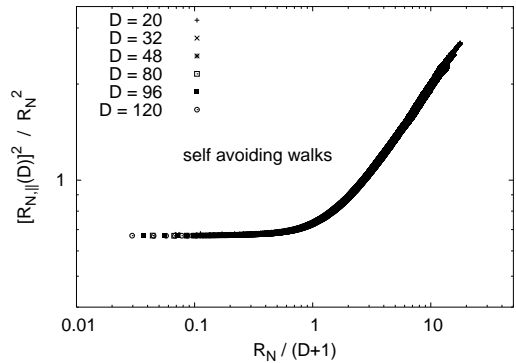,width=5.0cm, angle=270}
   \caption{Data collapse for testing the cross-over ansatz Eq.~(\ref{R2par})
     for self avoiding walks. }
    \label{saw-rr}
  \end{center}
\end{figure}

\begin{figure}
  \begin{center}
\psfig{file=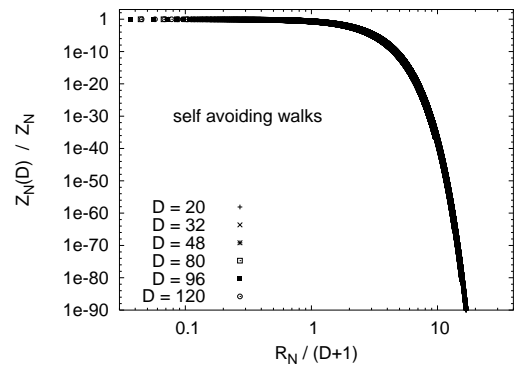,width=5.0cm, angle=270}
   \caption{Data collapse for testing the cross-over ansatz Eq.~(\ref{ZND1})
     for self avoiding walks. }
    \label{saw-zn}
  \end{center}
\end{figure}

In our non-trivial simulations we used widths up to $D=120$ and chain
length up to $N=80,000$. As first tests we checked the cross-over 
ansatzes Eqs.~(\ref{R2par}) and (\ref{ZND1}). In these tests we replaced
$R_N$ and $Z_N$ by parametrizations similar to Eqs.~(\ref{R2par}) and (\ref{ZND1}), 
but including additional correction to scaling terms. As seen from 
Figs.~\ref{saw-rr} and \ref{saw-zn}, the data collapse is excellent. Such a 
perfect collapse would not have been obtained, if we had replaced 
$R_N$ and $Z_N$ by the leading asymptotic powers of $N$~\cite{footnote}.

\begin{figure}
  \begin{center}
\psfig{file=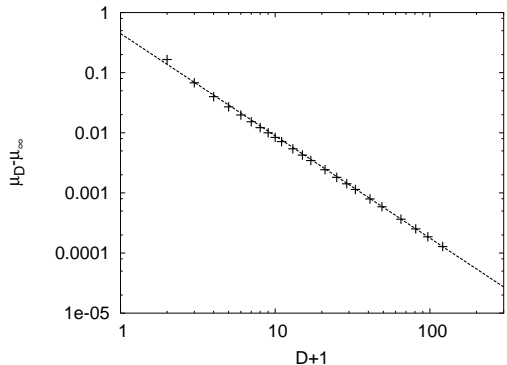,width=5.0cm, angle=270}
   \caption{Log-log plot of $\mu_D-\mu_\infty$ against $D+1$. The dashed line
            is $\mu_D-\mu_\infty = 0.448 (D+1)^{-1/\nu_3}$.}
    \label{saw-mu}
  \end{center}
\end{figure}

Critical fugacities were determined by plotting $\log Z_N -(\gamma_2-1)\log N
+ N x$ against $N$ and changing $x$ until these curves become horizontal for 
large $N$. Then $\mu(D) = \exp(x)$. Results are shown in Fig.~\ref{saw-mu},
where we plot $\mu(D)-\mu_\infty$ against $D$. As for ordinary RWs, the 
plot does not give a straight line (replacing $D$ by $D+1$ would improve the 
situation a bit, but not much), so the straight line shown in Fig.~\ref{saw-mu}
indicates the estimated asymptotic behaviour, assuming its slope to be 
given by $-1/\nu_3$. It provides us with the estimate $a=0.448\pm 0.005$.

\begin{figure}
  \begin{center}
\mbox{(a)}
\psfig{file=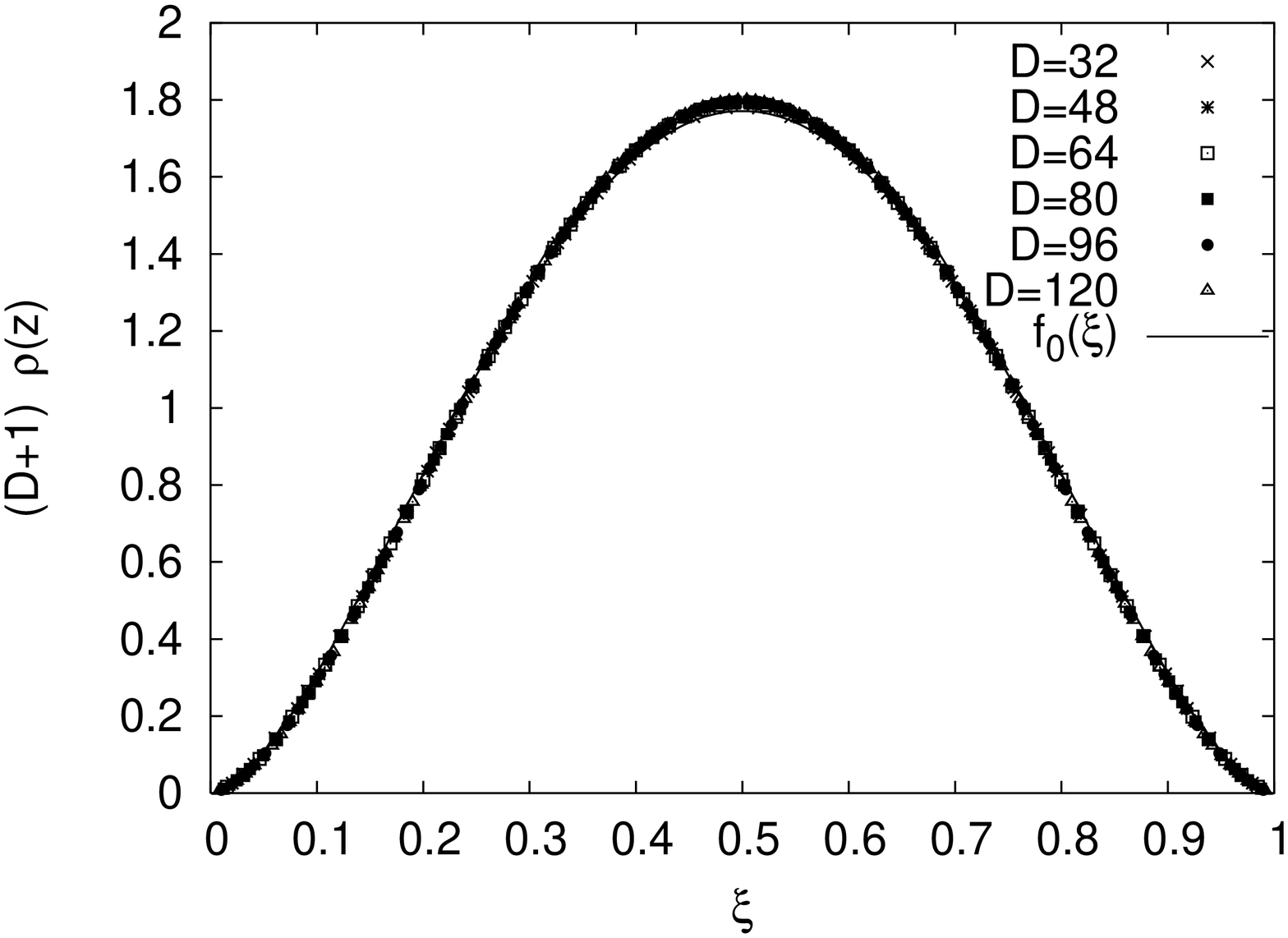,width=5.0cm, angle=270} \\
\mbox{(b)}
\psfig{file=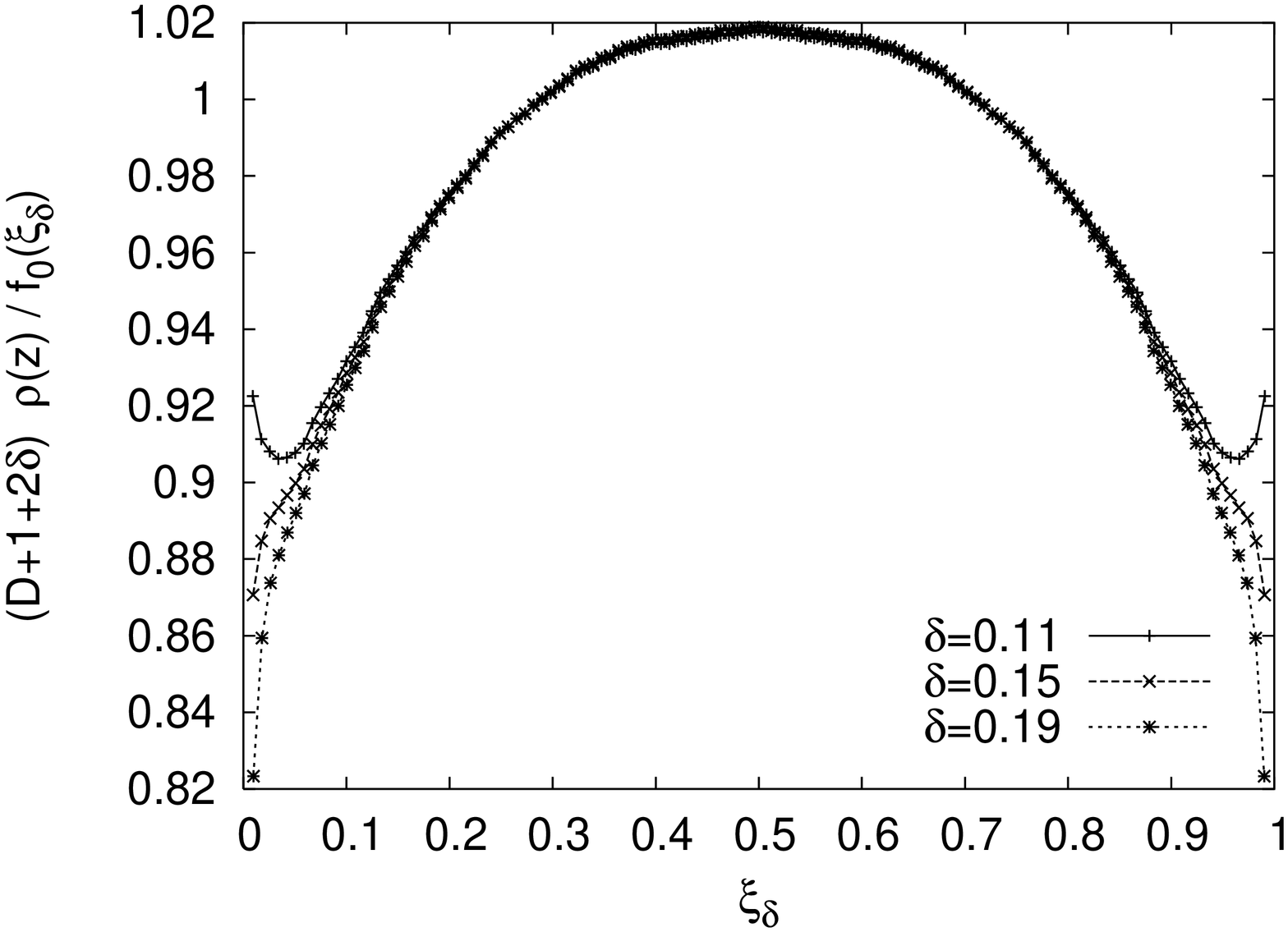,width=5.0cm, angle=270} \\
\mbox{(c)}
\psfig{file=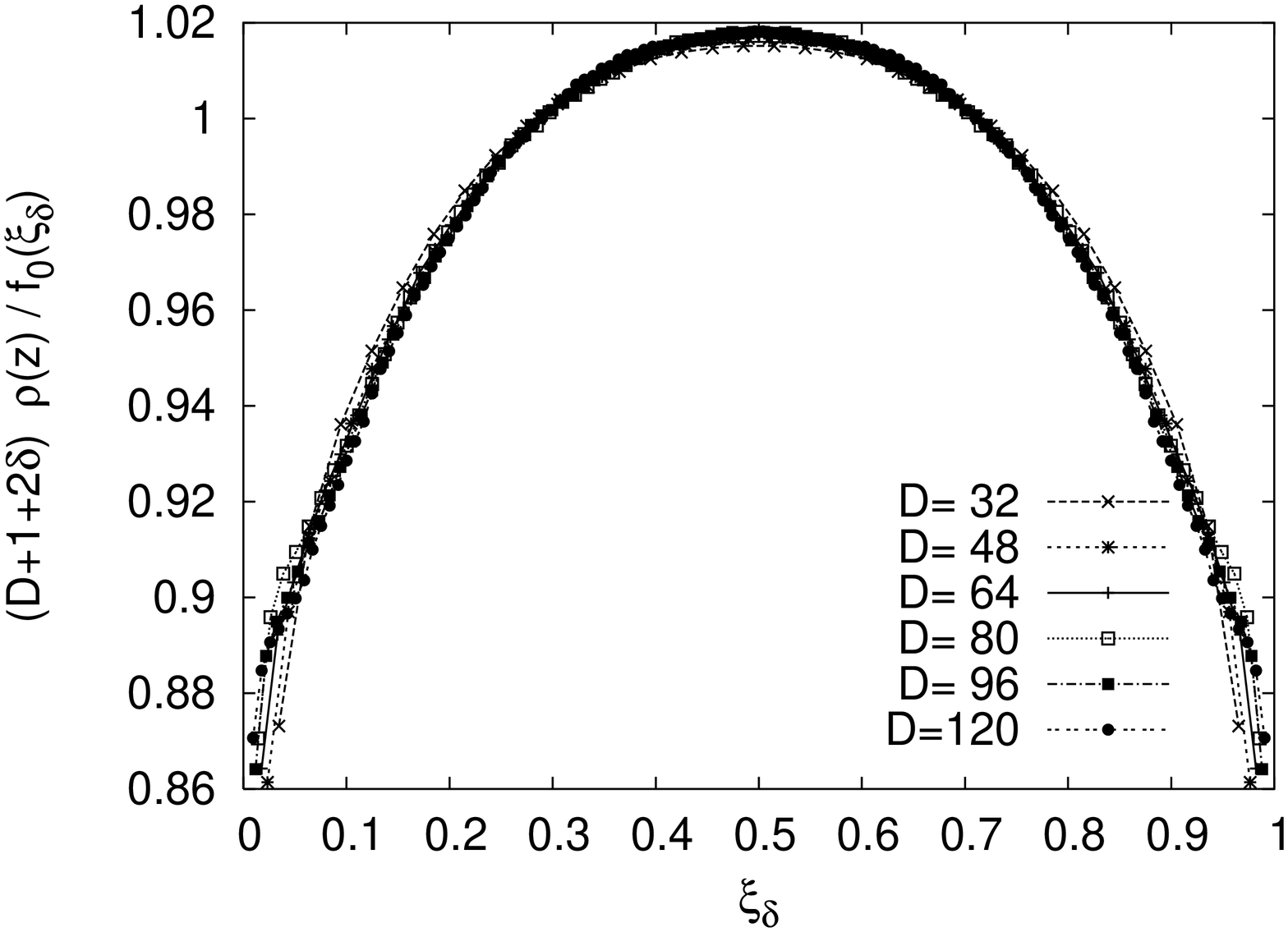,width=5.0cm, angle=270}
   \caption{(a) Rescaled values of the monomer density, $(D+1)\;
      \rho(z)$ against $\xi=z/(D+1)$. Also plotted is the
      function $f_0(\xi) = 18.74\;(\xi(1-\xi))^{1/\nu_3}$.\\
     (b) The data for $D=120$ plotted against a modified scaling variable,
       $\xi_\delta = (z+\delta)/(D+1+2\delta)$, and divided by
       $f_0(\xi_\delta)$, for three different values of $\delta.$\\
     (c) The data for all $D$, again divided by $f_0(\xi_\delta)$ with $\delta = 0.15$.}
    \label{saw-rhoz}
  \end{center}
\end{figure}

The monomer densities for different values of $D$, again from the central 
region only, are shown in Fig.~\ref{saw-rhoz}. Plotting the densities directly 
as in panel (a) indicates that scaling is satisfied. But it is not very 
informative, since deviations from scaling in the important regions near the 
walls would not show up. Also, panel (a) might suggest that $\rho(z)$ is simply
the product of two powers,
\be
    \rho(z) \approx {1\over D+1}f_0({z\over D+1}) \quad {\rm with}\quad f_0(\xi) = A\;
    [\xi(1-\xi)]^{1/\nu_3}\;,
\ee
where the constant $A=18.74$ is determined by normalization. We had already 
seen for random walks that $\rho(z)$ is not that simple, and indeed 
plotting $(D+1)\rho(z)/f_0(z/(D+1))$ as in panels (b) and (c) shows that 
this would be a very bad approximation. In addition, panel (b) shows
that one has to introduce an `extrapolation length' $\delta$ as suggested in
Ref.~\cite{Milchev}, 
so that the scaling variable $\xi$ is replaced by
\be
   \xi_\delta = {z+\delta\over D+1+2\delta}.
\ee
Best scaling near $z=0$ and $z=D$ (panel (b)) and best data collapse (panel 
(c)) is obtained for $\delta \approx 0.15$, although a closer inspection 
of these figures shows that neither the scaling nor the data collapse 
are perfect. These small persistent discrepancies and the overestimation 
of the amplitude $B$ discussed in the next paragraph were the main reason 
for studying the Domb-Joyce model.

Figure~\ref{saw-rhoz}c suggests that $D^{1+1/\nu_3} \rho(z)/z^{1/\nu_3} \to 
0.87(4)\times A = 16.1(8)$ for $z\to 0$ and $D\to\infty$. The very large 
uncertainty reflects the rather steep slopes at $z=0$ and $z=D$. Using this
in Eq.~(\ref{rhof}) gives $B=2.13\pm 0.11$.
This is larger than the prediction of Eisenriegler~\cite{Eisenriegler}, but
much less so than previous estimates~\cite{Milchev,Joannis}. We believe 
that these previous authors had missed the fact that $\rho(z)/f_0(\xi)$
is not constant. If we would assume $\rho(z)\propto f_0(\xi)$, we would 
obtain $B\approx 2.48$ which is indeed similar to the previous 
Monte Carlo estimates.

\begin{figure}
  \begin{center}
\psfig{file=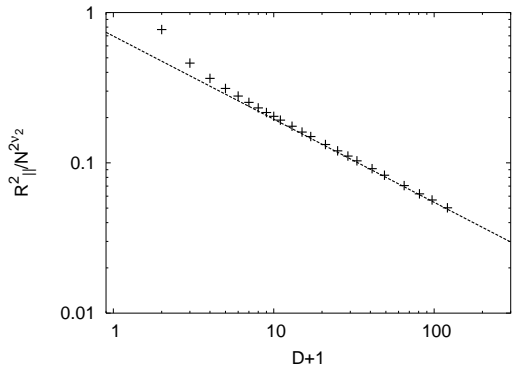,width=5.0cm, angle=270}
   \caption{log-log plots of $R^2_{||}(D)/N^{2\nu_2}$ versus $D$.
   The dashed line is $0.697\;(D+1)^{-0.553}$. }
    \label{saw-r2}
  \end{center}
\end{figure}

\begin{figure}
  \begin{center}
\psfig{file=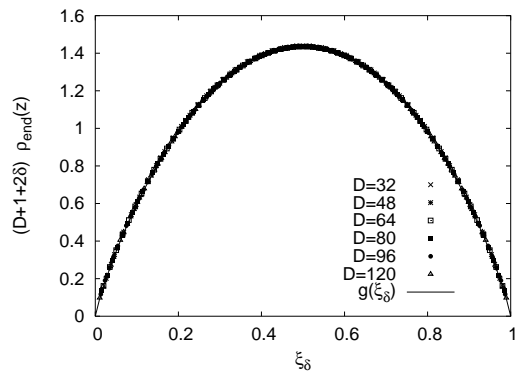,width=5.0cm, angle=270}
   \caption{Rescaled values of the probability $\rho_{end}(z)$ that the chain
       end is at the distance $z$ from a wall, against
       $\xi_\delta=(z+\delta)/(D+1+2\delta)$ with $\delta=0.3$.
       The solid line is the function $g(\xi_\delta)=4.78\;(\xi_\delta(1-\xi_\delta))^{0.865}$.}
            \label{saw-rhoend}
  \end{center}
\end{figure}

As a further test of scaling we checked in detail that $R_{||}(D) \sim N^{\nu_2}$
for $N^{\nu_2}\gg D$, and we estimated the asymptotic ratios between the two.
They are plotted in Fig.~\ref{saw-r2}, where we also plotted the scaling 
prediction,
\be
  R_{||}(D)^2/N^{2\nu_2} \sim D^{-2(\nu_2-\nu_3)/\nu_3} = D^{-0.553} \;.   \label{xn}
\ee 

Finally, we show in Fig.~\ref{saw-rhoend} the distribution $\rho_{end}(z)$ of chain 
ends. 
We found that $\rho_{end}(z)$ is very closely proportional to 
$(\xi_\delta(1-\xi_\delta))^{0.865}$ with $\delta=0.3$, but the (very small) 
deviations are highly significant. Taking them into account, we find
\be
   \rho_{end}(z) \sim z^{0.80(2)}
\ee
near the walls, with $\delta\approx 0.2$. This agrees nicely with Eq.~(\ref{rhoend}).

\subsection{Domb-Joyce model}

Domb-Joyce chains with interaction strength $w=0.6$, which is very close to 
critical strength $w^*$ where leading corrections to scaling 
vanish~\cite{gss,belohorec}, were studied for slab widths up to $D=80$. Chain 
lengths were up to $N=72000$. The analysis of the data was done exactly as 
for the self avoiding walks described in Sec. IIIB.

\begin{figure}
  \begin{center}
\psfig{file=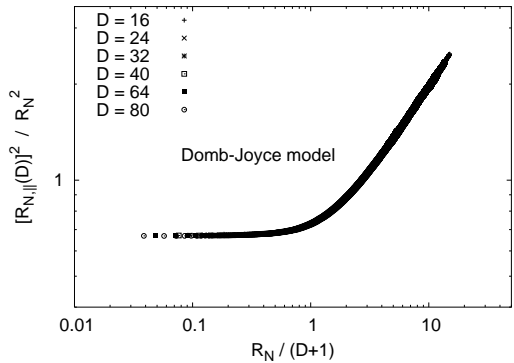,width=5.0cm, angle=270}
   \caption{Data collapse for testing the cross-over ansatz Eq.~(\ref{R2par})
     for Domb-Joyce walks. }
    \label{dj-rr}
  \end{center}
\end{figure}

\begin{figure}
  \begin{center}
\psfig{file=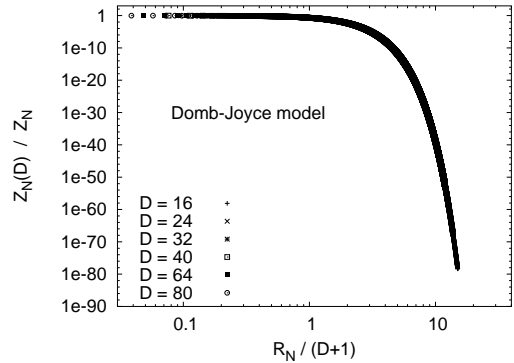,width=5.0cm, angle=270}
   \caption{Data collapse for testing the cross-over ansatz Eq.~(\ref{ZND1})
     for Domb-Joyce walks. }
    \label{dj-zn}
  \end{center}
\end{figure}

As expected, the scaling functions $\Phi(\eta)$ and $\Psi(\eta)$ are very
similar to those for SAWs (see Figs.~\ref{dj-rr} and \ref{dj-zn}). 
This universality verifies that the amplitudes and critical exponents 
discussed in Sec.~II are essentially correct, although this should not 
be taken too serious: Such data collapse plots are not very sensitive to
details (look at the huge range of scales in Fig.~\ref{dj-zn}!).

\begin{figure}
  \begin{center}
\psfig{file=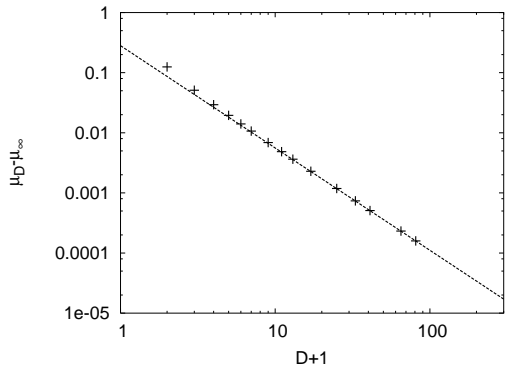,width=5.0cm, angle=270}
   \caption{Log-log plot of $\mu_D-\mu_\infty$ against $D+1$ for the Domb-Joyce
       model. The dashed line
       is $\mu_D-\mu_\infty = 0.2813 (D+1)^{-1/\nu_3}$ with $\nu_3$ used as
       constraint.}
    \label{dj-mu}
  \end{center}
\end{figure}

Estimates of the critical fugacities are shown in Fig.~\ref{dj-mu},
where we plot $\mu_D-\mu_\infty$ against $D+1$. The straight line, which again
represents the extrapolation to large $D$, provides the estimate $a=0.2813(6)$.

\begin{figure}
  \begin{center}
\mbox{(a)}
\psfig{file=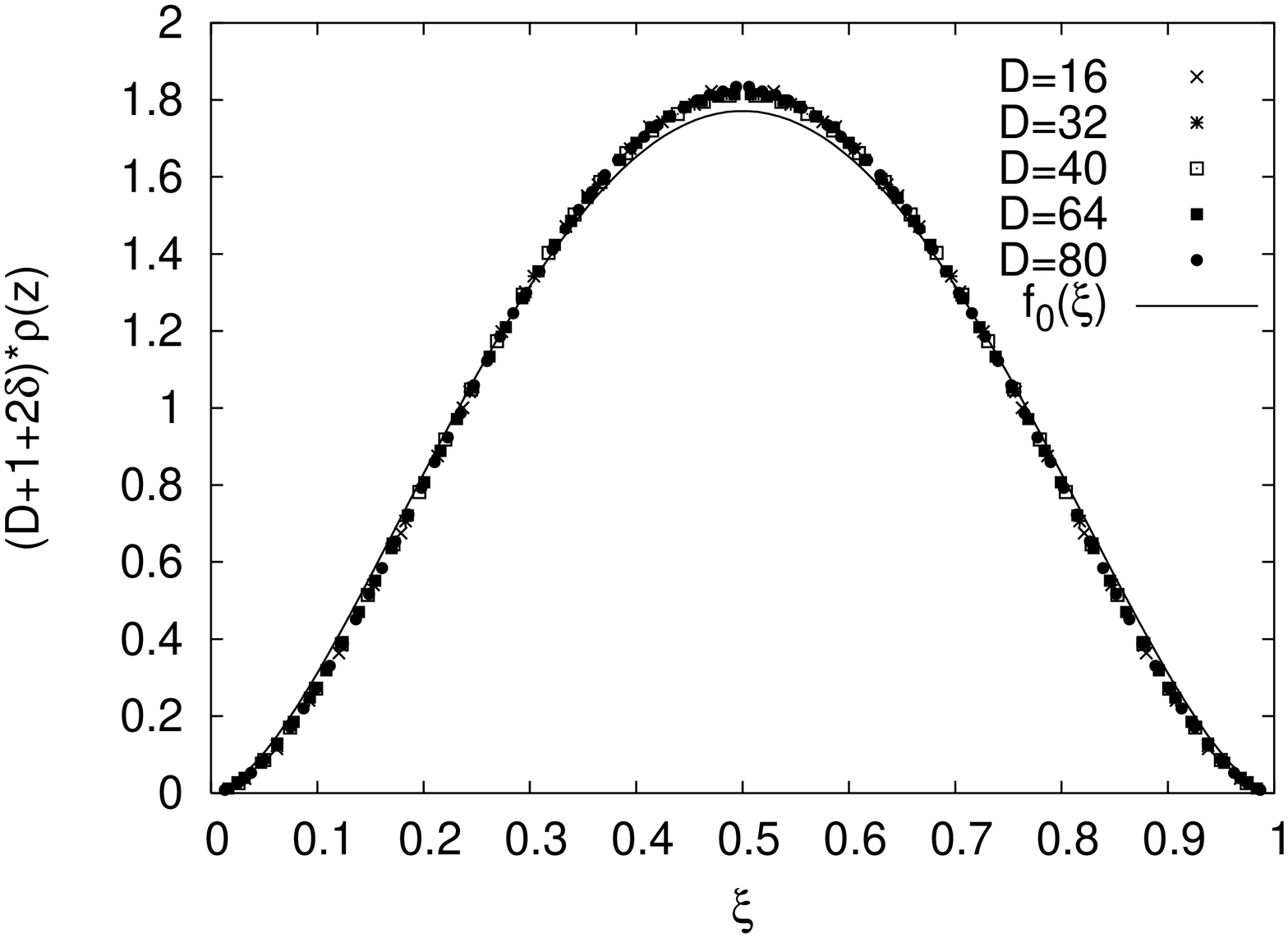,width=5.0cm, angle=270} \\
\mbox{(b)}
\psfig{file=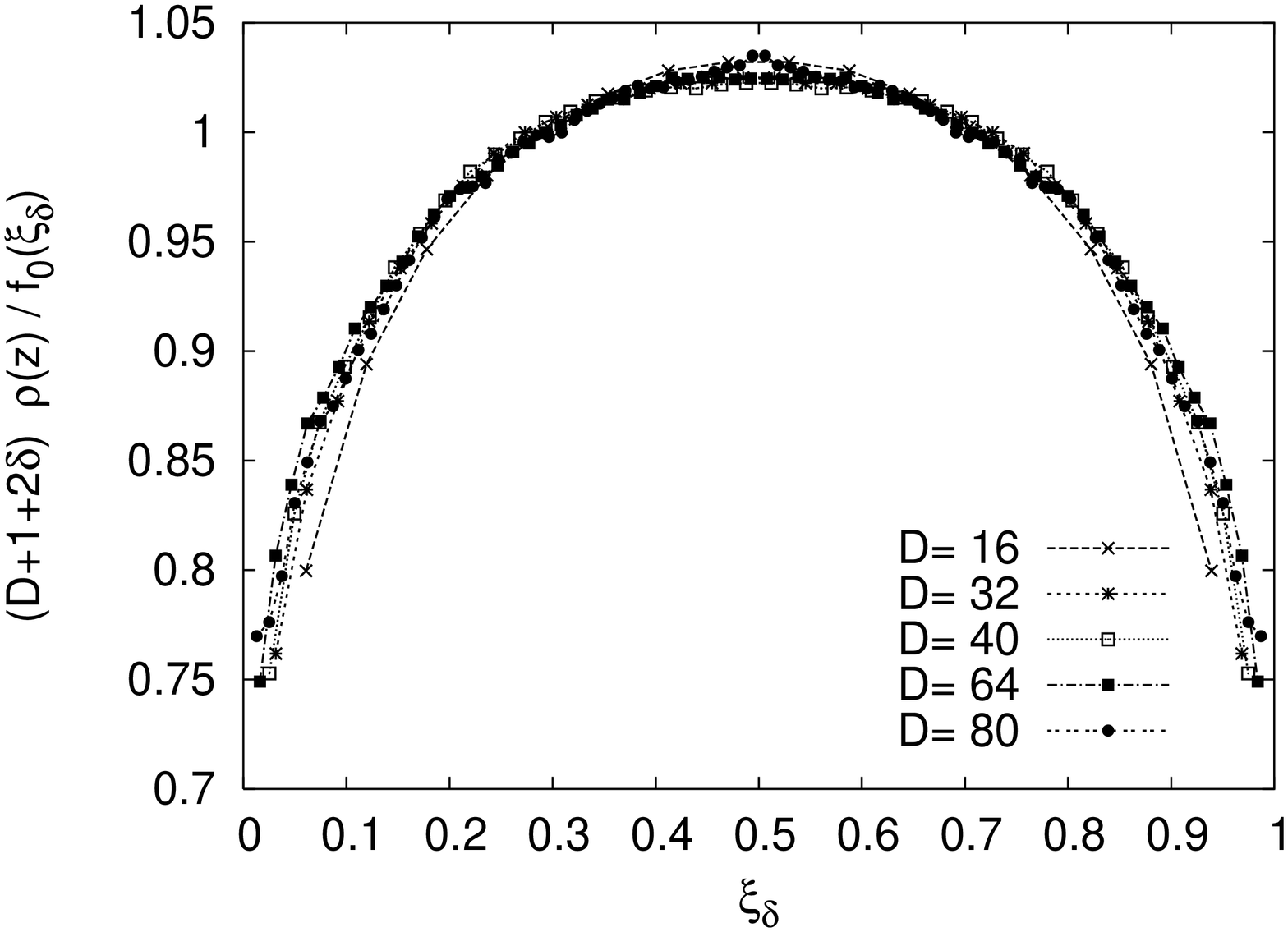,width=5.0cm, angle=270}
   \caption{(a) Rescaled values of the monomer density $(D+1+2\delta)\;
      \rho(z)$ of the Domb-Joyce model
       against $\xi=(z+\delta)/(D+1+2\delta)$ with $\delta=0.06$.
      Also plotted is the function $f_0(\xi) = 18.74\;(\xi(1-\xi))^{1/\nu_3}$.
     (b) The same values as in panel (a), but divided by $f_0(\xi_\delta)$ with
       $\delta=0.04$.}
    \label{dj-rhoz}
  \end{center}
\end{figure}

Plots of the monomer density profile (Fig.~\ref{dj-rhoz}) are very similar 
to those for SAWs. But the extrapolation length is now much smaller, 
$\delta\approx 0.04$ as compared to $\delta\approx 0.15$ for SAWs. This 
is a first indication that corrections to scaling are indeed smaller in
the DJ model. More important, also the scaling curve in Fig.~\ref{dj-rhoz}c
looks slightly different from that in Fig.~\ref{saw-rhoz}b: It is considerably 
smaller at the walls, with $\lim_{z\to 0,\;D\to\infty} D^{1+1/\nu_3}z^{-1/\nu_3}\rho(z)/A = 0.71(3)$
as compared to 0.87(4) for SAWs. Given the fact that scaling corrections should 
be smaller for the DJ model (in spite of the somewhat smaller values of $D$),
we consider the DJ value as more correct, and blame the discrepancy onto
scaling corrections for SAWs. With this new estimate of 
$\lim_{z\to 0,\;D\to\infty} D\rho(z)$, and using the non-universal amplitude
$a$ determined earlier, we obtain our final estimate for the universal
amplitude ratio $B$,
\be
   B = 1.70\pm 0.08.
\ee
This is only 2 standard deviations away from the $\epsilon$-prediction $B=1.85$ of
Eisenriegler~\cite{Eisenriegler}, which we consider as good agreement.

\begin{figure}
  \begin{center}
\psfig{file=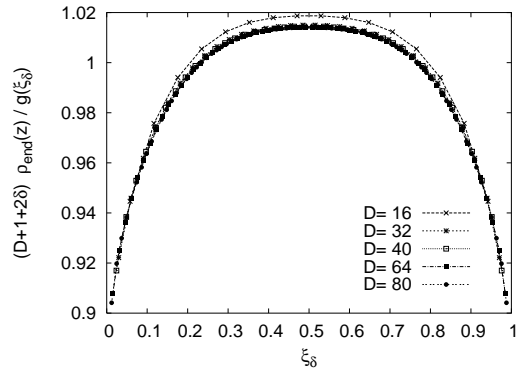,width=5.0cm, angle=270}
 \caption{Rescaled values of the probability $\rho_{end}(z)$ that the chain
      end is at the distance $z$ from a wall, divided by the function
      $g(\xi_\delta)=4.358(\xi_\delta(1-\xi_\delta)^{0.81}$, against
      $\xi_\delta=(z+\delta)/(D+1+2\delta)$ with $\delta=-0.02$.}
        \label{dj-rhoend}
  \end{center}
\end{figure}

Finally, we do not show our data for $R_{N,\|}(D)/N^{\nu_2}$ and for the 
end monomer profile, since they are very similar to Figs.~\ref{saw-r2} and 
\ref{saw-rhoend}. But again the seemingly perfect agreement of the end
point density profile is again as deceptive as it was for SAWs. This time
our best estimate for the scaling of the end point distribution is
\be
   \rho_{end}(z) \sim z^{0.81(1)}\;,
\ee
with $\delta \approx -0.02$, in even better agreement with Eq.~(\ref{rhoend})
than the estimate for SAWs. The rescaled density profile divided by
$g(\xi_\delta) = 4.358(\xi_\delta(1-\xi_\delta))^{0.81}$ is shown in 
Fig.~\ref{dj-rhoend}.

\section{Summary}

We have presented high statistics simulations of 3-d polymers, modelled as 
walks on a simple cubic lattice with either hard or soft excluded volume 
interactions, with chain length up to 80000 on slabs of widths up to 120. This 
was possible with the PERM algorithm with Markovian anticipation. The fact
that PERM gives by default very precise estimates of free energies allowed
us to measure precisely the forces exerted onto the walls, by measuring how 
the critical fugacities depend on the width of the slabs. We verified all
critical scaling laws predicted for this problem, including the scaling of 
monomer and end point densities near the walls and the scaling of the total
pressure with chain length and with slab width.

The theoretical prediction most difficult to verify numerically concerns 
the amplitude ratio between the pressure onto the wall and the monomer 
density close to the wall. Previous simulations had not been able to obtain
this with sufficient precision, and also in the present paper we had serious
problems when using self avoiding walks with strict (hard-core) self 
repulsion. This might not be so surprising, given the well known fact that
SAWs show rather large corrections to scaling. These corrections to scaling
can be minimized by going over to Domb-Joyce polymers (characterized by soft 
repulsion) with carefully adjusted strength of the repulsion (similarly, for
off-lattice bead-spring models, one can adjust the ratio between bead size
and equilibrium spring length to minimize corrections to scaling). It was 
only when going over to this Domb-Joyce model, that we could verify in 
detail all theoretical predictions.

Thus we have shown, first of all, that already the field theoretic 
$\epsilon$-expansion to first order in $\epsilon$, as implemented 
in Ref.~\cite{Eisenriegler},
gives correct results. This was not obvious, in particular
in view of persistent previous difficulties to verify it by Monte Carlo 
simulations. Secondly, we have demonstrated again the importance of using 
models with minimized corrections to scaling. And last not least we have
again shown that recursive sequential sampling methods with
re-sampling~\cite{liu} (of which PERM is a particular implementation) can be very 
efficient.

Acknowledgements: We thank Profs.~Erich Eisenriegler and Ted Burkhardt for 
valuable discussions, and Dr.~Walter Nadler for carefully reading the manuscript.


\begin{thebibliography}{99}
\bibitem{footnote} While $R^2_{N,\|}(D)$ increases monotonically with $1/D$,
    this is not true for the full 3-dimensional $R^2_N(D)$, since the 
    compression in $z$-direction overwhelms the swelling in $x$ and $y$, 
    for large and intermediate values of $D$~\cite{vliet,cordeiro}.
\end{thebibliography}
\end{document}